\documentclass[aps,pra,amsfonts,floatfix,superscriptaddress,twocolumn,longbibliography,nofootinbib]{revtex4-1}

\newcommand{\comment}[1]{}
\usepackage{graphicx}% Include figure files
\usepackage{dcolumn}% Align table columns on decimal point
\usepackage{bm}% bold math
\usepackage{amsmath}
\usepackage{amssymb}
\usepackage{bbold}
\usepackage{braket}
\usepackage{physics}
\usepackage{float}
\usepackage{mathtools}
\usepackage{multirow}
\usepackage{tikz}
\usepackage{adjustbox}
\usepackage{soul,color}
\usepackage{xcolor}
\usepackage{hyperref}
\usepackage{algorithm}
\usepackage{multirow}
\usepackage{algpseudocode}
\usepackage{MnSymbol}
\begin{document}
\preprint{APS/123-QED}
%\title{Online Learning of Random Unitary Channels with Contracted\\Quantum Approaches and Simplex Optimization}
% Scott's Original: Learning Random Unitary Channels Gradually with Contracted Quantum Approaches and Simplex Optimization
\title{Learning Orthogonal Random Unitary Channels with Contracted Quantum Approaches and Simplex Optimization}

\author{Scott E. Smart}
\thanks{These authors contributed equally}
\email{sesmart@ucla.edu}
\affiliation{College of Letters and Science, University of California, Los Angeles, California 90095}
\author{Alexander J{\"u}rgens}
\thanks{These authors contributed equally}
\affiliation{Dept. of Electrical \& Computer Engineering, University of California, Los Angeles, California 90095}
\affiliation{Institute for Theoretical Physics, ETH Z{\"u}rich, 8093 Zurich, Switzerland}

\author{Joseph Peetz}	
\affiliation{Department of Physics and Astronomy, University of California, Los Angeles, California 90095}
\author{Prineha Narang}
\email{prineha@ucla.edu}
\affiliation{College of Letters and Science, University of California, Los Angeles, California 90095}
\affiliation{Dept. of Electrical \& Computer Engineering, University of California, Los Angeles, California 90095}

\date{\today}

\newcommand{\E}{\mathcal{E}}
\newcommand{\F}{\mathcal{F}}
\begin{abstract}
Random (mixed) unitary channels describe an important subset of quantum channels, which are commonly used in quantum information, noise modeling, and quantum error mitigation. Despite their usefulness, there is substantial complexity in characterizing or identifying generic random unitary channels. We present a procedure for learning a class of random unitary channels on orthogonal unitary bases on a quantum computer utilizing Pauli learning and a contracted quantum learning procedure. Our approach involves a multi-objective, Pauli- and unitary-based minimization, and allows for learning locally equivalent channels. We demonstrate our approach for varying degrees of noise and investigate the scalability of these approaches, particularly with sparse noise models.   
\end{abstract}

\maketitle

\section{Introduction}\label{section: intro}

The characterization of quantum channels plays a crucial role in designing and realizing quantum technologies \cite{Bialczak_2010,Tinkey_2021,PhysRevLett.97.220407,peters2005precisecreationcharacterizationmanipulation}. These are described by completely positive trace preserving (CPTP) superoperators, capturing both their internal unitary dynamics and interaction with the outside world \cite{nchuang,MRenes+2022}. Quantum process tomography (QPT), the process of completely determining a given quantum channel through state preparation, channel application and measurement, is a task that is as crucial and ubiquitous as it is challenging. Due to the exponentially increasing dimension of the Hilbert space and superoperator space, as well as potential numerical sensitivity, QPT quickly becomes intractable. Additionally, only very few protocols exist that are able to introduce bias in the tomography procedure to learn a quantum channel of a fixed non-general form. % These are the issues we are seeking to tackle in this work. 

A particularly interesting subset of CPTP maps are random unitary channels (RUCs), the convex hull of the unitary channels. RUCs are of frequent use in noise modeling and device characterization \cite{Emerson_2005}. They contain common error channels, including dephasing, depolarizing, and coherent errors \cite{Audenaert2007, watrousMixingDoublyStochastic2009}, as well as a closed ball around the completely depolarizing channel \cite{watrous2008mixingdoublystochasticquantum}, while still being efficiently simulated on quantum devices \cite{scottpeetz}. They are also of interest from an error correction perspective as they exactly describe errors that can be corrected using information from the environment \cite{Gregoratti_2003}, and also have been utilized in error mitigation procedures \cite{kimEvidenceUtilityQuantum2023}. RUCs are related to mixed states \cite{RUC_Complementarity}, as any mixed state can be characterized as an RUC applied to a set reference \cite{brunsWitnessingRandomUnitary2016b}. 

Despite their usefulness, the problem of learning generic RUCs is particularly challenging. Multi-qubit RUCs are not unique, and there exists no canonical form. Just discerning whether a given unital channel is an RUC or not is NP-hard \cite{leeDetectingMixedunitaryQuantum2020}. Several recent works have explored learning a subset of RUCs, Pauli channels, demonstrating an exponential speed-up over naïve channel tomography \cite{flammiaEfficientEstimationPauli2020,chenQuantumAdvantagesPauli2022a,nchuang}. Without entanglement, sparse polynomially scaling Pauli channels can be learned \cite{vandenbergProbabilisticErrorCancellation2023}. Within the context of gate set tomography \cite{nielsenGateSetTomography2021, briegerCompressiveGateSet2023a}(where unitary errors can be accounted for by calibration), a mixed RUC with measurement errors was investigated \cite{moueddeneRealisticSimulationQuantum2020}. A recent approach for small unitary and Pauli errors was introduced, though requiring measurement of the exponentially scaling Pauli transfer matrix \cite{kaufmannCharacterizationCoherentErrors2023}. Regardless, both of these methods have similar elements to costs and concerns for generic QPT, and do not allow for learning generic RUC. 

In this work, we consider a biased tomography approach for learning RUCs, and specifically ones that admit an orthogonal representation under the Hilbert Schmidt inner product, which we refer to as ORUCs. Our online learning approach is inspired by work in iterative state tomography \cite{bandit}, and involves alternating updates between a Pauli channel and input and output unitary channels. For the unitary learning procedure, we introduce a contracted quantum approach \cite{smartQuantumSolverContracted2021}, requiring very few quantum evaluations. We investigate a sparse additive Pauli model with least-squares learning procedure as well as a Riemannian optimization of the probability simplex manifold, the latter of which again allows for updates with a single Pauli circuit measurement. We show the sparse Pauli and unitary frameworks are scalable, and can be used together in a multi-objective framework for learning RUCs.  We also discuss and highlight some of the limitations of these models and evaluate the performance on examples beyond the model, particularly in finding non-orthogonal random unitaries and non-equivalent error bases.

\section{Theoretical Motivation}

Let $\mathcal{H}$ be a Hilbert space of dimension $d = 2^n$. The set of all Pauli operators on $n$ qubits is denoted as $\mathbb{P}^n$, with elements $\sigma_k$ and $|\mathbb{P}^n|=4^n$. Channel composition between two channels $\mathcal{E}$ and $\mathcal{F}$ is denoted by $\mathcal{E} \circ \mathcal{F}$. Given a channel $\E$, we can express it in a Pauli transfer matrix $T_{\alpha \beta} = {\rm Tr} \sigma_\alpha \E[\sigma_\beta] = \llangle \sigma_\alpha | \E | \sigma_\beta \rrangle$. Throughout the work, we use the Pauli basis for input states and so commonly describe $\rho_\alpha$ as a Pauli matrix with unit norm (i.e. $\rho_\alpha = \frac{1}{d}\sigma_\alpha$). The inner product here is the Hilbert-Schmidt inner product. A Haar-random unitary refers to a randomly selected (according to the Haar measure) element of $\mathcal{U}(d)$ \cite{Bengtsson_Zyczkowski_2006}, distinguished from a random unitary channel. Local equivalence refers to unitaries acting of a bipartite subsystem whereas a $k-$local unitary refers to an operator acting on $k-$subsystems.

A quantum channel $\mathcal{E}$ is called a random unitary channel (or a mixed unitary channel) if it can be written in the form:
\begin{equation}
    \mathcal{E}[\rho] = \sum_i p_i U_i^{} \rho U_i^\dagger   
\end{equation}
where $U_i$ are a collection of unitary operators, and $p_i$ forms a convex set of coefficients. The smallest number of terms $N$ denotes the random unitary rank, which can be larger than $d$ but is bounded by the Choi rank $r$ ($N \leq r^2 -r +1$) \cite{girardMixedunitaryRankQuantum2022b}. Given a unital channel, the problem of discerning whether or not the channel admits a random unitary form is NP-hard \cite{leeDetectingMixedunitaryQuantum2020}.  

If we restrict ourselves to the case where the set of $\{U_i \}$ are orthogonal under the Hilbert-Schmidt norm, i.e., ${\rm Tr}~U_i^\dagger U_j = \delta_{ij} d $, then we have at most $d^2$ elements and if so, can describe the set of unitaries as a unitary basis \cite{schwingerUnitaryOperatorBases1960, howeNiceErrorBases2005, knillGroupRepresentationsError1996a}. The most common basis within quantum information is the Pauli basis, although the Weyl basis, or ``nice'' bases with group-theoretic structure are common as well \cite{howeNiceErrorBases2005,klappeneckerMonomialityNiceError2005} 

Two unitary bases $\{ T_i\} $ and $\{ S_i\}$ are said to be locally unitarily equivalent \cite{klappeneckerMonomialityNiceError2005} if there exist some unitary transforms $U$ and $V$, such that:
\begin{equation}
T_i = U S_i V 
\end{equation}
for all elements of the unitary basis. Unitary bases also correspond to orthogonal sets of maximally entangled states through vectorization, and these local unitary transformations are requisite to preserve the maximally entangled nature of the states \cite{poonPreserversMaximallyEntangled2015,brunsWitnessingRandomUnitary2016b}.

We refer to random unitary channels where the $U_i$ are elements of an orthogonal basis as orthogonal random unitary channels (ORUCs). ORUCs are a subset of all possible random unitary channels, with maximal random unitary rank $d^2$. For single qubit channels, all unital channels are locally equivalent to some Pauli channel \cite{Audenaert2007}
For more than one qubit, it is possible to generate unital transforms that are not orthogonal random unitary \cite{watrousUnitalChannelsMajorization2018}, as well as non-locally equivalent random unitary channels \cite{klappeneckerMonomialityNiceError2005} with differing error bases. For the present work, we focus on learning random unitary channels in the equivalence class of the Pauli channel:
\begin{equation}\label{eq:sqoruc}
    \mathcal{E}[\rho] = \mathcal{E}_U \circ \mathcal{E}_P \circ \mathcal{E}_V[\rho] = \sum_i p_i U \sigma_i V \rho V^\dagger \sigma_i U^\dagger,
\end{equation} 
noting that other nice error bases could also be used. %An alternative form of equivalence, setting $V = U^\dagger W$, provides an other useful framework for multi-objective optimization, but is not completely compatible with some of the approaches. 

For $n$ qubits, there are $4^n$ Pauli basis elements, and the two unitaries can be globally parameterized by $2\cdot{4^{n}}-2$ parameters. While this is much less complex than attempting to parameterize a general random unitary channel, this still presents a problem with exponential scaling. However, for most applications we have some form of structure present, and so for many applications will look at random unitaries with rank $p << 4^d$, or that are products of smaller unitaries. While the methods are not restricted to sparse channels, these allow for more efficient learning to take place, and we treat the optimization procedure with stochastic methods due to the high dimensionality of the parameter space and stochastic updates. The approach below has the advantage of being theoretically applicable to sparse and non-sparse channels, though we expect the rate of convergence to be proportional to the sparsity. 

%Additionally, the ideas here can be applied broadly to generic random unitary channels, but removing the condition of local equivalence which relaxes the condition on the error basis. Given a random unitary channel $\mathcal{E}$, we denote the rank of the Choi–Jamiolkowski isomorphism as $R_\mathcal{E}$. Then, there exists a random unitary form with at most $R_\mathcal{E}^2$ operators. Using this, we can \emph{naively} parameterize a vector of $\mathbb{R}^2$, in addition to a list of  unitaries (of dimension $d^2$) for each vector. This is not ideal computationally, hence the current work focuses on orthogonal random unitary channels. 

\section{Online Orthogonal Random Unitary Learning Procedure}
We introduce an iterative learning approach inspired by gradient-descent state tomography. Direct process tomography is unable to discern between unital and random unitary forms, and variational approaches that incorporate random unitary forms often suffer from high gradient costs and issues related to the high dimensionality of parameterization. The works in Refs.~ \cite{moueddeneRealisticSimulationQuantum2020, kaufmannCharacterizationCoherentErrors2023} use simultaneous learning procedures for their models, but have limited generalizability.

%and so we treat the process tomography procedure as a learning procedure, where we start with the true channel $\mathcal{E}_0$ and an estimate ORUC $\mathcal{E}_g$, and iteratively improve our estimate.

Let $\mathcal{F}$ denote the channel of interest (which may not even be unital) and $\mathcal{E}$ the trial channel in Eq.~\eqref{eq:sqoruc}. Then, we define a loss function 
with respect to randomly sampled measurements $\sigma_\alpha$ and input states $\rho_\beta$:
\begin{equation}
\mathcal{L}(\sigma_\alpha,\rho_\beta) = \frac{1}{2}|\llangle \sigma_\alpha | \E - \F | \rho_\beta \rrangle|^2.
\end{equation}
The Pauli channel learning focuses on $\alpha=\beta$, whereas the unitary learning exclusively uses $\alpha \neq \beta$. While there are numerous approaches for both, we focus on near-term learning, with the overarching multi-objective optimization described in Algorithm~\ref{alg:gen}. We further detail the selection of $\sigma_\alpha$ and $\rho_\beta$ in Appendix~\ref{sec:computational}.

Generically, we denote a multi-stage optimization as one where the relevant criteria can change. This allows for dynamic weighting of different variables, though is not unique in how we treat multiple objectives. Additionally, because the learning procedures can be cast as single input/observable estimations, we can remove statistical uncertainty by sampling more points.  

\begin{algorithm}[H]
\caption{Alternating multi-objective learning procedure for ORUC. Given: a target channel $\mathcal{E}$, learning methods $\textsc{learn}_\mathcal{P}$ and $\textsc{learn}_\mathcal{U}$ (defined in Algorithms \ref{alg:learnp} and \ref{alg:learnu}) learning the Pauli channel and the unitaries $U,V$ for $N_k^U$ and $N_k^p$ sub-iterations respectively, a distance measure $d$ in superoperator space, maximal number of iterations $K$ and termination precision $\epsilon$. Outputs ORUC approximation $\hat\E$ to target channel.}\label{alg:gen}
\begin{algorithmic}[1]
\State \textbf{inputs}: $\E$, $K$, $d$, $\epsilon$ $N_k^U$, $N_k^p$, $\textsc{learn}_\mathcal{P}$, $\textsc{learn}_\mathcal{U}$, optional $\{\E_{U}, \E_V, \E_P\}$
\State \textbf{initialize}: $\hat\E \gets \E_{U} \circ \E_{P} \circ \E_V$
\State $k \gets 0 $
\While{$k < K$ or $d(\E,\hat \E)\geq\epsilon_k$}\Comment{}
\State $\E_U,\E_V \gets \textsc{learn}_\mathcal{U}(\mathcal{E},\E_{U}, \E_{V}, \E_P, N^U_k) $
\State $\E_P \gets  \textsc{learn}_\mathcal{P}({\E_{U^{-1}}\circ \E\circ\E_{V^{-1}}}, \E_P,N^p_k) $
\State $\hat\E \gets \E_{U} \circ \E_{P} \circ \E_V$
\State $k \gets k+1$
\EndWhile\label{euclidendwhile}
\State \textbf{return} $\hat\E$
\end{algorithmic}
\end{algorithm}

\subsection{Learning Unitary Channels}
Unitary channel learning has many forms, depending on access to unitaries as oracles or quantum memory \cite{angrisaniLearningUnitariesQuantum2023,xueVariationalQuantumProcess2022,zhaoLearningQuantumStates2024, bisioOptimalQuantumLearning2010}. In the context of near-term quantum channel learning, variational approaches are common \cite{galetskyOptimalDepthNovel2024, xueVariationalQuantumProcess2022}, particularly in machine learning applications \cite{jonesRobustQuantumCompilation2022b, anschuetzQuantumVariationalAlgorithms2022a}. 

\subsubsection{Variational Unitary Ansatz}
Given parameterizable unitaries $U(\theta)$ and $V(\phi)$, we substitute the unitary learning problem with a unitary parameterization problem. Essentially, we take a simple variational unitary ansatz, and then can define expectation values:
\begin{equation}
x_\alpha(\theta,\phi) = {\rm Tr} ~ \sigma_\alpha U(\theta) \mathcal{E}_P[ V(\phi) \rho_\alpha V^\dagger(\phi)] U^\dagger(\theta)
\end{equation}
and a canonical loss function
\begin{align}\label{eq:loss}
\begin{split}
    \mathcal{L} = \sum_{\alpha} \frac{1}{m} \mathcal{L}_\alpha = \frac{1}{2m}\sum_{\alpha=1}^m (y_\alpha -x_\alpha)^2
    \end{split}
\end{align}
here $m$ denotes the number of measurement observables, and may or may not include the dimension of the system. For simple types of parameterizable unitaries, we can derive gradient rules using the parameter shift rule \cite{crooksGradientsParameterizedQuantum2019}, resulting in gradients measured simply from the loss function evaluated at different parameters. This implies a simple learning procedure for $\theta$ and $\phi$:
\begin{equation}
\frac{d}{d \theta_i} \mathcal{L} = \sum_\alpha \frac{1}{m}(x_\alpha - y_\alpha) x_\alpha'
\end{equation}
and $x'_\alpha$ is measured from the parameter shift rules. Parameters are then updated by a learning rate $\mu_\mathcal{U}$. The parameterization of the unitaries in general is not trivial, as well as efficiently measuring the gradient terms, which is proportional to the number of parameters.

\subsubsection{Contracted Quantum Learning}

Here we introduce a contracted quantum learning approach, inspired by the contracted quantum eigensolver from many-body electronic structure theory \cite{smartQuantumSolverContracted2021}. We consider two infinitesimal unitary channels generated by anti-Hermitian matrices $A$ and $B$, which modify the input and output local unitary channels respectively. At each step, we seek to find a direction of $A$ and $B$ which minimizes our loss function. The total procedure is specified in Algorithm \ref{alg:learnu}. 

The gradient with respect to these operators represents the contracted equation (see Appendix \ref{sec:con_lstsq}), which can also be realized by considering variations of the loss function. This relates somewhat to layerwise learning \cite{skolikLayerwiseLearningQuantum2021}, although the choice of problem, optimization, and execution differ substantially. The loss parametrized by a small time step $t$ is given as:
\begin{equation}
x_\alpha(t) = {\rm Tr} \left[ \sigma_\alpha e^{tB} \E[e^{t A} \rho_\alpha  e^{-t A}] e^{-t B} \right].
\end{equation}
Decomposing $A$ and $B$ into the Pauli basis (i.e. $ A = \sum_k i a_k \sigma_k$), we can describe the derivatives with respect to these coefficients at $t=0$:
\begin{equation}
\frac{\partial }{\partial a_k} x_\alpha(t) |_{t=0} =  i{\rm Tr} \sigma_\alpha \mathcal{E}\big[[\sigma_k,\rho_\alpha]\big],
\end{equation}
\begin{equation}
\frac{\partial }{\partial b_k} x_\alpha(t) |_{t=0} =  i{\rm Tr} [\sigma_\alpha, \sigma_k] \mathcal{E}[\rho_\alpha].
\end{equation} 
When input into the derivative of the loss, in the CQE literature these are equivalent to the (anti-Hermitian portion of the) contracted equations \cite{mazziottiAntiHermitianPartContracted2007}, although these are not defined with respect to a Hamiltonian. While similar, these are not exactly the same as the direct minimization of the skew-symmetric Lie algebra (see \cite{smartManybodyEigenstatesQuantum2024a} for related discussion). The commutator terms are either $0$ or another Pauli string, indicating that for each pair of inputs and outputs $(\sigma_\alpha, \rho_\alpha)$, the gradient with respect to the input and output coefficients are given by circuits over $(\sigma_\alpha, \rho'_\alpha)$ and $(\sigma'_\alpha, \rho_\alpha)$ respectively. 

Letting $A'$ and $B'$ simply represent the minus gradient direction (i.e., $a'_k = - [y_\alpha - x_\alpha(0)]\frac{\partial}{\partial a_k}x(t)|_{t=0}$) and choosing a learning rate $\eta$, we form the corresponding unitary channels:
\begin{equation}
\mathcal{E}_{\exp \eta A'}[\rho] = e^{\eta A'} \rho ~e^{-\eta A'}, 
\end{equation}
with the following updates:
\begin{align}
\mathcal{E}_V &\leftarrow \mathcal{E}_V \circ \mathcal{E}_{\exp \eta A'} \\
\mathcal{E}_U &\leftarrow \mathcal{E}_{\exp \eta B'} \circ \mathcal{E}_U
\end{align}

In contrast with the variational parameterization above, we have several key advantages. The first is that the local minima necessarily correspond to a solution of the contracted equation (see Appendix \ref{sec:con_lstsq}). This implies that we are at a stationary point with respect to all local unitary operators (or with respect to our generating set), which can be a stronger condition than the parameterized quantum circuit. The second is that the derivatives for a reduced set of generators can often be measured more efficiently. For instance, the set of all two-local qubit operators, $\sigma_\alpha = \sigma_k^i \sigma_k^j$ when combined with two-local observables, can only generate up to 3-local terms (i.e. $[X_1 X_2, Z_2 Z_3] = -2i X_1 Y_2 Z_3$). Accordingly, these can be measured relatively efficiently via local grouping of qubits \cite{Bonet-Monroig2019}. However, the iterative nature of the ansatz presents challenges. For small dimensions, we take advantage of circuit compilation to construct a compact unitary that does not grow with respect to system size. For one- and two-qubit unitaries this can be done exactly, though in general there are also heuristics that exist allowing for compact ansatz to be formed \cite{smartAcceleratedConvergenceContracted2022}.  

\begin{algorithm}[H]
\caption{Generic contracted quantum learing algorithm for use in algorithm \ref{alg:gen} for learning the outer and inner unitary channels $U,V$. Given: A target channel $\F$, learning rate $\eta$, number of iterations $N^U$, current guesses $U,V$, indices $\mathcal{I}_\mathcal{A}$ of a set of generators $\mathcal{A}$, and Pauli channel $\E_P$}\label{alg:learnu}
\begin{algorithmic}[1]
\State \textbf{inputs}: $\F,\mu,N^U,\E_U,\E_V,\E_P$
\State \textbf{initialize}: $\E\gets\E_U\circ\E_P\circ\E_V$
\State \textbf{sample}: $[\sigma_{\alpha_1},...,\sigma_{\alpha_{N^U}}], [\rho_{\beta_1},...,\rho_{N^U}] \subset\mathbb P^n, \mathbb P^n$
%\State \textbf{sample}: $[\sigma_1,...,\sigma_{N^U}]\subset\mathbb P^n$
%\State \textbf{sample}: $[\rho_1,...,\rho_{N^U}]$
\For{{$i=1,...,N^U$}}
    \State \textbf{measure} $y_\alpha=\text{Tr}[\sigma_{\alpha_i}\F(\rho_{\beta_i})]$
    \State \textbf{measure} $x_\alpha=\text{Tr}[\sigma_{\alpha_i}\E(\rho_{\beta_i})]$
    \For{{$k \in \mathcal{I}_\mathcal{A}$}}
        \State \textbf{measure} $\frac{\partial }{\partial a_k} x_{\alpha_i}(t) |_{t=0} =  i{\rm Tr} \sigma_{\alpha_i} \mathcal{E}\big[[\sigma_k,\rho_{\beta_i}]\big]$
        \State \textbf{measure} $\frac{\partial }{\partial b_k} x_{\beta_i}(t) |_{t=0} =  i{\rm Tr} [\sigma_{\alpha_i}, \sigma_k] \mathcal{E}[\rho_{\beta_i}]$
    \EndFor
    \State \textbf{initialize} $A=-\sum_k i[y_{\alpha_i} - x_{\alpha_i}]\frac{\partial}{\partial a_k}x(t)|_{t=0}~\sigma_k$
    \State \textbf{initialize} $B=-\sum_k i[y_{\beta_i} - x_{\beta_i}]\frac{\partial}{\partial b_k}x(t)|_{t=0}~\sigma_k$
    \State $\mathcal{E}_V \gets \mathcal{E}_V \circ \mathcal{E}_{\exp \eta A}$
    \State $\mathcal{E}_U \gets \mathcal{E}_{\exp \eta B} \circ \mathcal{E}_U$

\EndFor
\State \textbf{return} $\mathcal{E}_U,\mathcal{E}_V$
\end{algorithmic}
\end{algorithm}
\subsubsection{Reducing Measurements Resources through Resolution of the Identity}\label{sec:ri}
The online-learning procedure can be further modified via a relation over the channel trace distance (see Appendix \ref{sec:app_loss}). This allows us to isolate the contracted equations via two modified loss functions. Namely, at each step, we apply the adjoint of our channel to the actual channel and construct two modified loss functions:
\begin{align}
\mathcal{L}_A &= \frac{1}{2}|\llangle \sigma_\alpha | \E_{\exp \mu A} - \E_V^\dagger  \E_U^\dagger \F | \rho_\alpha\rrangle |^2  \\
\mathcal{L}_B &= \frac{1}{2}|\llangle \sigma_\alpha | \E_{\exp  \mu B} - \F \E_V^\dagger \E_U^\dagger | \rho_\beta \rrangle|^2 
\end{align}
Under an average over all input states and measurement bases, these loss functions share the same stationary condition as the target channel and the guess channel Eq.~\eqref{eq:sqoruc}. However, these have the distinction that the derivative expression is
\begin{equation}
\frac{d}{d a_k} \llangle  \sigma_\alpha | \E_{\exp \mu A}  | \rho_\beta \rrangle|_{\mu=0} = 2i \delta^{\alpha}_{[k,\beta]}
\end{equation}
where $[k,\beta]$ is the Pauli matrix element under the commutator, potentially zero, and can be evaluated classically. Thus, the total number of function evaluations, regardless of number of parameters, is 2. The classical overhead can vary, but this {substantially} reduces the quantum resources. After the direction is calculated, the unitaries $U$ and $V$ can be updated and their inverse incorporated into the channel as well. We summarize these methods in Table~\ref{tab:compare}. 

\begin{table}

\renewcommand{\arraystretch}{1.25}
\caption{Quantum circuit evaluations for function and gradient evaluations of the variational unitary learning procedures.}

\begin{tabular}{c|ccc }
\multirow{2}{*}{Method}&\multicolumn{2}{c}{Quantum Calls} & \multirow{2}{*}{Fixed Circuit}  \\ 
 & Function & Gradient & \\ \hline
PQC ($U$, $V$) & $2$ & $2N_p^U + 2N_p^V$ & Yes \\
CQL & $2$ & $\mathcal{O}([\sigma_k,\sigma_\alpha])$ & No \\
RI CQL  & $2$ & $2$ & No
\end{tabular}
\label{tab:compare}
\end{table}

\begin{figure}
\centering
\vspace{-0.25cm}
\includegraphics[scale=1]{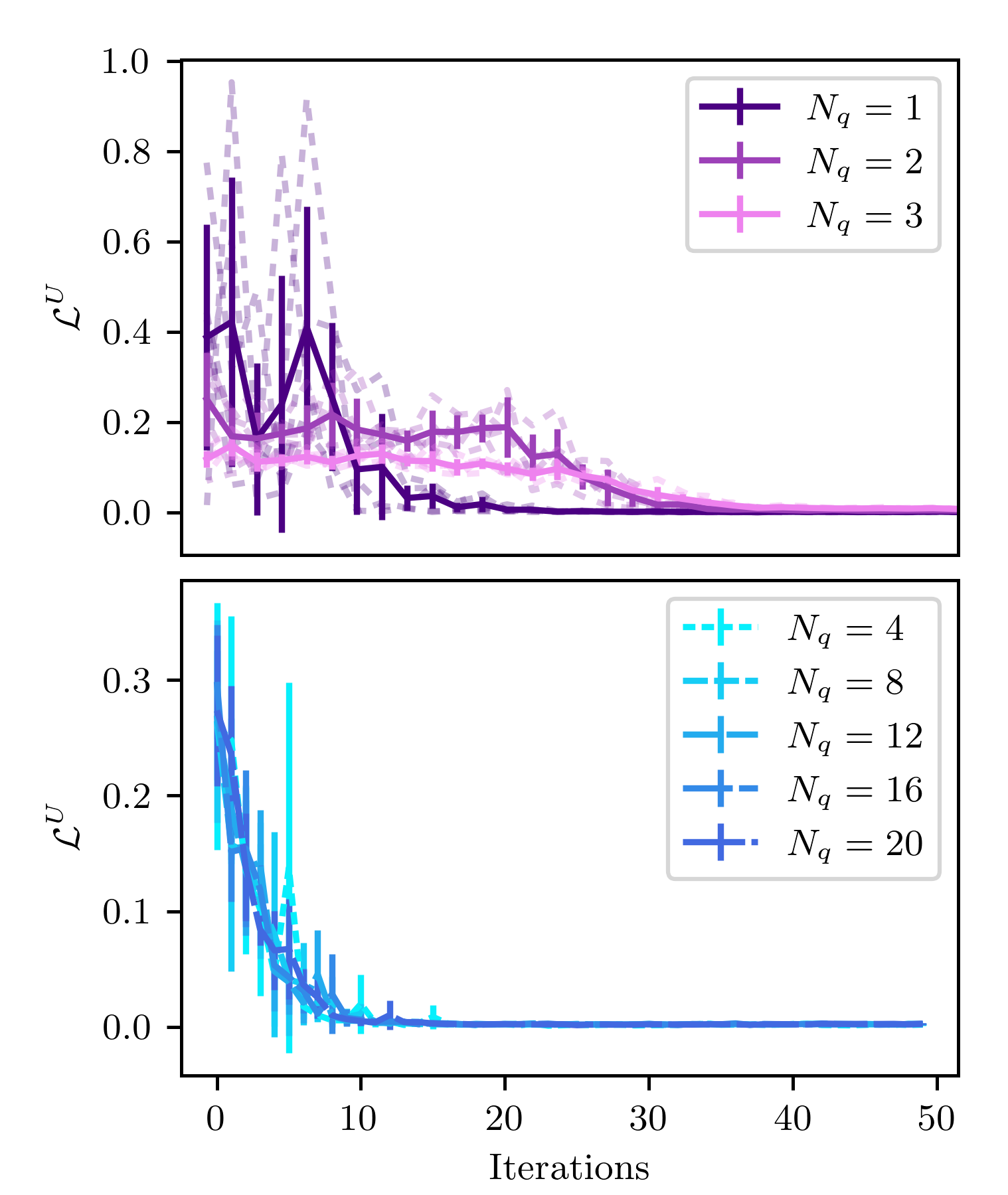}
\caption{(Top) Contracted quantum learning procedure for multi-qubit Haar random unitaries. (Bottom) Scaling of independent single qubit channel. Results are over five runs, and the loss is normalized with respect to the number of qubits or the degree of generators. Each iteration uses a single sampled set of Pauli operators measured simultaneously. }
\label{fig:learn_U}
\end{figure}

We demonstrate the contracted quantum learning procedure for orthogonal random unitary channels in Figure~\ref{fig:learn_U} for multiple qubit channels, as well as its scalability for local, independent channels, namely an independent system of $1-$local unitary errors. Independent one- and two-qubit channels are highly scalable, as these possess exact transpilation procedures to concatenate a series of unitary transformations. However, Haar random unitaries are notoriously difficult to learn and generally require an exponential parameterization. Despite this, we do not see a substantial increase in the number of iterations required to obtain higher unitaries, though we expect to explore this in future work.  

\subsection{Learning Pauli Channels} We can write any channel acting on a state in the Pauli transfer representation \cite{greenbaumIntroductionQuantumGate2015}:
\begin{align}
\mathcal{E}_{ref}[\rho] = \sum_{\alpha \beta }\llangle \alpha | \E_{ref} | \beta \rrangle \llangle \beta | \rho \rrangle .
\end{align} 
Pauli channel learning typically involves learning the diagonal coefficients, corresponding to the twirled channel. These can be transformed via the Walsh-Hadamard transform to the coefficients of the corresponding Pauli channel \cite{caiConstructingSmallerPauli2019}. 

\subsubsection{Current Approaches}

An efficient means of learning a Pauli channel was presented by Wallman \& Flammia \cite{wallmanEfficientLearningPauli2019} as well as Chen et al. \cite{chenQuantumAdvantagesPauli2022a}. The central idea is to prepare a maximally mixed state and then measure in the Bell basis. The measurement outcomes over $2 n$ qubits express the symplectic representation of the Pauli strings and thus directly outputs the probability of measuring a Pauli fidelity. We can directly use this by inserting the inverse unitary channels into the circuit before and after the application of the channel, and then measuring it. Note that this gives the effective twirled Pauli channel, with a corresponding vector $\vec{p}_0$. 

A particular challenge for near-term computing is separating the effects of the channel from the Bell-state preparation. Even mild errors can substantially bias and change the output probability distribution, resulting in the inclusion of terms at a much higher proportionality. While the number of qubits can be reduced, this is with an exponential trade off in the number of required circuits.  Additionally, for many physical systems, we expect that operator locality plays a key role, and many physical systems might have noise with only $1-$ or $2-$local Pauli strings.

A recent result also introduced a sparse Pauli-Lindblad model \cite{vandenbergProbabilisticErrorCancellation2023}, where instead of learning a single Pauli channel, the authors learn a composition of Pauli channels, which commute and can be multiplicatively implemented. This method allows for many simple random unitary channels to be learned, which then are combined. These approaches may also be used for our procedure, though may not be as efficient for the iterative procedure. 

\subsubsection{Via Linear Inversion}

We first introduce a loss function in the Pauli channel basis. That is, we evaluate:
\begin{align}
\vec{y}_\alpha &= {\rm Tr}~\sigma_\alpha \mathcal{E}_0 [\rho_\alpha] \\
\vec{x}_\alpha =  {\rm Tr}~\sigma_\alpha \mathcal{E}_P &[\rho_\alpha] =  \sum_k p_k (-1)^{\langle \sigma_\alpha, \sigma_k \rangle } \delta_{\sigma_\alpha}^{\rho_\alpha}
\end{align}
where $\langle \cdot,\cdot \rangle$ is the symplectic inner product of the two Pauli strings indicating their commutativity and $p_k$ represents elements of a probability vector $\vec{p}$. Letting $S$ be a matrix with elements denoting the inner product, $S_{\alpha \beta} = (-1)^{\langle \sigma_\alpha, \sigma_\beta \rangle }$, we can define the generic loss function in a vectorized form:
\begin{equation}
\vec{\mathcal{L}} = \frac{1}{2}(\vec{y} - S \vec{p})^T(\vec{y} - S \vec{p})
\end{equation}
taking the derivative with respect to $\vec{p}$ yields a unique solution:
\begin{equation}
\vec{p} = S^{-1} \vec{y}.
\end{equation}

As long as we choose elements of our set of Paulis such that the matrix $S$ is invertible, we can solve for $\vec{p}$, though we do not have guarantees that the resulting vector is positive or normalized. Thus, we include an update procedure which performs the following:
\begin{align}
\vec{p}  &\leftarrow  P_{\mathcal{S}^d}\big((1-\mu)\vec{p} + \mu S^{-1} \vec{y}\big)
\end{align}
where $P_{\mathcal{S}^d}(\cdot)$ is a projection operator onto the appropriate $d-$dimensional simplex and $\mu$ is a learning rate. This serves to zero out negative values and renormalizes the state.

\subsubsection{Via Riemannian Gradient Descent }

\begin{figure}[t]
\includegraphics[scale=1.0,trim={0 0 0 0.5cm}, clip]{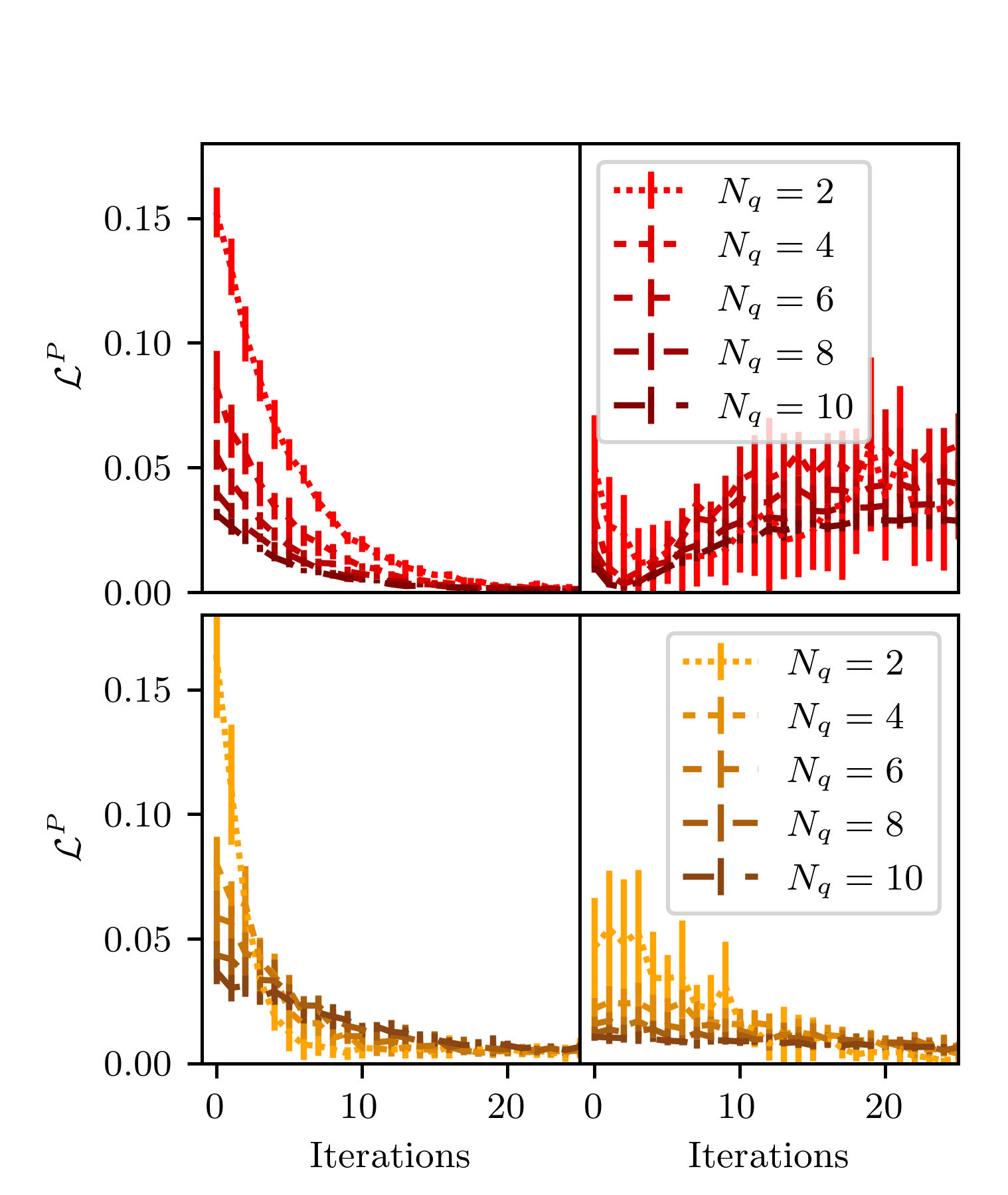}
\vspace{-0.5cm}
\caption{Learning single-qubit Pauli noise models with increasing system size, where target $p_I \approx 0.6$ in an additive Pauli channel, averaged over 10 random sets of Paulis. (Top) Least-squares approach using a simple inversion of the symplectic matrix $S$.  (Bottom) Riemannian optimization of the simplex with learning rate $0.75$. (Left) Full Pauli set of expectation values. (Right) Learning with single-Pauli expectations.}
\label{fig:fig_multi_pauli}
\vspace{-0.5cm}
\end{figure}

We can also treat the interior of the probability simplex as a Riemannian manifold, and then substitute our current approach with classical Riemanian gradient descent \cite{coopeEfficientCalculationRegular2017, boumalIntroductionOptimizationSmooth2023, astromImageLabelingAssignment2017}. For points on the exterior, these can be treated effectively on the interior with a small $\epsilon$.  

Let $\mathcal{S}^d$ denote the $d-$dimensional probability simplex, and $\mathbb{1}^d$ denote a $d-$dimensional vector of ones. The tangent space of $\mathcal{S}^d$ at $\vec{p}$ is defined as $\mathcal{T}_{\vec{p}} = \{\vec{v} \in \mathbb{R}^d : \langle \mathbb{1}^d, \vec{v}\rangle  = 0 \}$. For our retraction, we simply use the projection of the first-order update onto the interior of the simplex. 

The Riemannian gradient for a smooth function $f$ from $\mathcal{S}^d$ to $\mathbb{R}$ is given by \cite{astromImageLabelingAssignment2017}:
\begin{equation}
{\rm \grad} f(\vec{p}) = \vec{p} \odot (\nabla f (\vec{p}) - \langle \vec{p}, \nabla f(\vec{p})\rangle \mathbb{1}^d). 
\end{equation} 
where $\odot$ indicates the Hadamard product. For the least squares problem with $\nabla f(\vec{p}) = S^T ( \vec{y} - S \vec{p})$, this results in the form:
\begin{align}
[{\rm \grad} ~\vec{\mathcal{L}}(\vec{p})]_k &= p_k ( g_k - \langle p, g \rangle )  \\
\implies{\rm \grad} ~\vec{\mathcal{L}}(\vec{p}) &= \vec p \odot( \vec g - \langle p, g \rangle \mathbb{1}^d )  
\end{align}
 where $\vec g=S^T ( \vec{y} - S \vec{p})$. Because $\vec{p}$ is sparse, this can be carried out completely classically. Furthermore, at each step ${\rm dim}(\vec{y})$ can be less than $d$, i.e. a single Pauli element, and we simply take limited slices of $S$. This allows us to easily make updates on $\vec{p}$ using limited expectation values, which is detailed in Algorithm \ref{alg:learnp}.

We compare the least-squares and simplex approaches in Figure~\ref{fig:fig_multi_pauli}, with both showing rapid convergence of the Pauli channels for the full set of expectations (left). While theoretically, these are similar approaches, for single-Pauli evaluations (right) the Riemannian optimization can reliably obtain the proper channel, in contrast with the least-squares approach, even with low learning rates. One strategy to address this would be with an iterative least-squares approach, where prior steps help constitute the vector $\vec{x}_\alpha$, although the multi-objective framework complicates this. Note we use an additive Pauli model here, in contrast to learning separate smaller channels (see Appendix \ref{sec:add}).

\begin{algorithm}[H]
\caption{Riemannian gradient descent based Pauli tomography algorithm for use in Algorithm \ref{alg:gen} for learning the locally equivalent Pauli channel. Given: a target Pauli channel $\E_0$ and a current estimate $\E_P$ given through $\vec p$, a leaning rate $\eta$ over a number of iterations $N^P$ and measurement batch size $N^B$.}\label{alg:learnp}
\begin{algorithmic}[1]
\State \textbf{inputs}: $\E_0,\vec p,\mu,N^P$
\State \textbf{sample}: $[(\sigma_{\alpha_1},\rho_{\alpha_1}),...,(\sigma_{\alpha_{N^B}},\rho_{\alpha_{N^B}})]\subset\mathbb (P^n,P^n)$
\For{{$i=1,...,N^B$}}
    \State \textbf{Measure} $y_i=\text{Tr}\big[\sigma_{\alpha_i}\E_0(\rho_{\alpha_i})\big]$
\EndFor
\State \textbf{initialize} $S\gets S_{\alpha \beta} = \left((-1)^{\langle \sigma_{\alpha_i}, \sigma_\beta \rangle }\right)_{i,\beta}$ \textbf{for $\beta\in P^n$}
\For{{$j=1,...,N^P$}}
\State \textbf{initialize} $\vec g\gets S^T ( \vec{y} - S p)$
\State \textbf{initialize} $\nabla \mathcal{L}\gets \vec p \odot( \vec g - \langle \vec p, \vec g \rangle \mathbb{1}^d )$
\State $\vec p \gets P_{\mathcal{S}_d}\left(\vec{p} - \eta \nabla \mathcal{L}\right)$
\EndFor

\State \textbf{return} $\E_P(\vec p)$
\end{algorithmic}
\end{algorithm}

\subsection{Learning Orthogonal Random Unitary Channels}
Learning the random unitary channel requires a composite learning procedure. 
Because the Pauli and unitary channels are not independent, the optimization is particularly challenging. The work in Ref.~ \cite{kaufmannCharacterizationCoherentErrors2023} presents one instance where a unified framework is possible, although the scope is limited to small angle unitaries and requires a large number of circuits. In general, we would like to properly incorporate knowledge of the Pauli channel into our unitary learning, and vice versa.

\begin{figure*}[t]
\centering
\includegraphics[scale=1.0, trim = {0.5cm 0 0  0cm}]{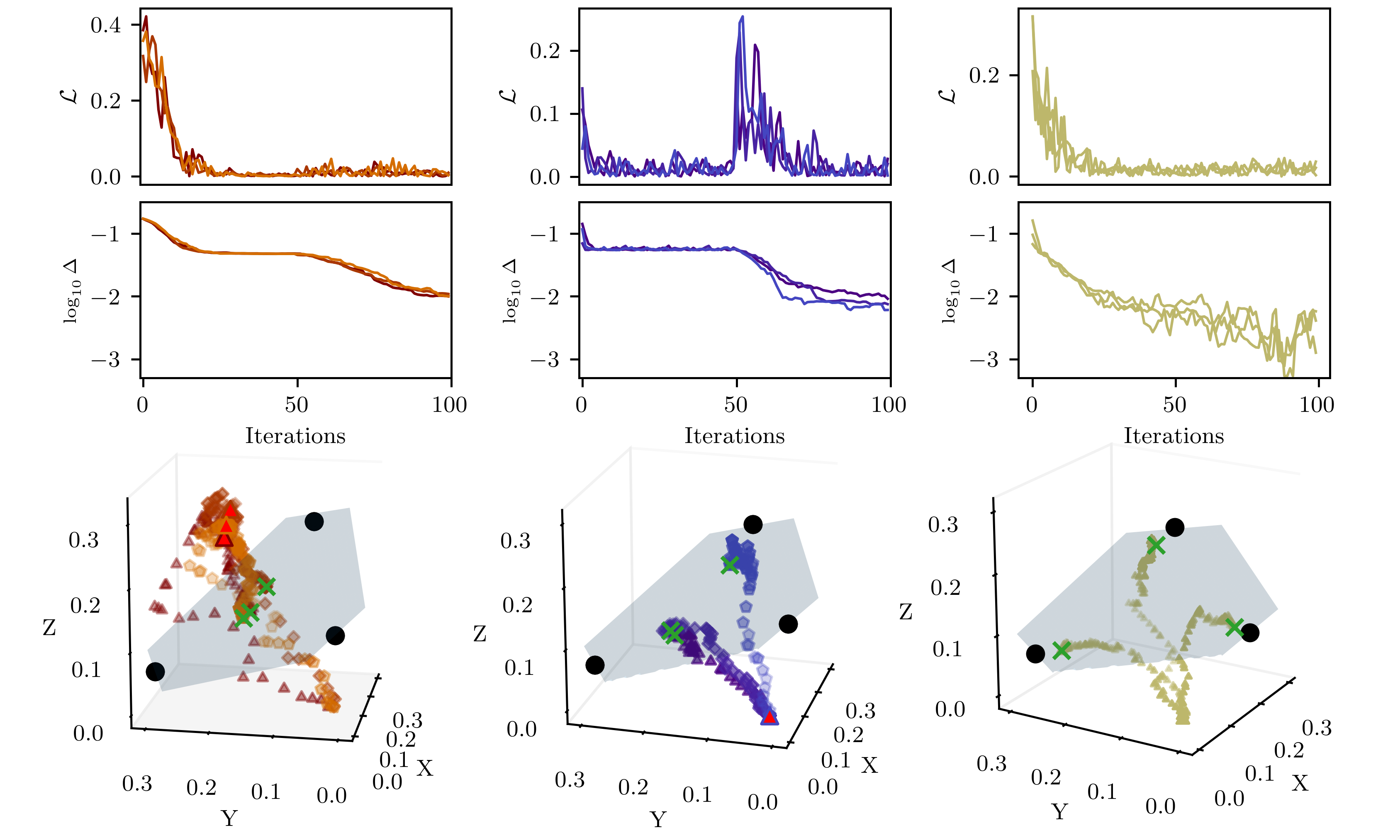}
\vspace{-0.5cm}
\caption{Convergence of ORUC learning procedure for three repetitions of different optimizations. The target channel here has coefficients $p_x = p_z = 0.05$, and $p_y=0.3$, with $U_f = \exp \frac{i}{\sqrt{4}}\sigma_x$ and $V_f = \exp -\frac{i}{\sqrt{4}}\sigma_z$. (Top) Pauli and Unitary loss functions with the lower box denoting the calculated total channel loss from the Pauli transfer matrices. (Bottom) Trajectories of the Pauli learning coefficients. The gray plane defined by $\sum p_k = 1- p_I$. Red ($\triangle$) represent the mid point, and green ($\times$) the final channel coefficients. Black circuits represent local unitarily equivalent distributions. (Left) Pauli$\rightarrow$Unitary learning procedure. (Middle) Unitary$\rightarrow$Pauli.  (Right) Simultaneous Pauli$\leftrightarrow$Unitary learning strategies.   More details are included in Section~\ref{sec:results}.
}
\label{fig:two_stage}
\end{figure*}

For learning the Pauli channel, we can simply apply the inverse unitary channels before and after the true channel. Similar to the method described in Section~\ref{sec:ri}, this loss function is equivalent to the standard loss. The expectations from the true channel can be obtained as:
\begin{equation}
\vec{y}_\alpha = {\Tr }~\sigma_\alpha \mathcal{E}_{U}^{-1} \circ \mathcal{E} \circ \mathcal{E}_V^{-1} [\rho_\alpha]
\end{equation}
whereas the expectations for the test channel are:
\begin{align}
\vec{x}_\alpha &=  {\rm Tr}~\sigma_\alpha \mathcal{E}_P[\rho_\alpha] =  \sum_k p_k (-1)^{\langle \sigma_\alpha, \sigma_k \rangle } \delta_{\sigma_\alpha}^{\rho_\alpha}
\end{align}
the latter of which can be calculated classically from $\vec{p}$.

For the unitary channel, we sample the Pauli channel with unitary transforms applied before and after. These expressions are described in Table \ref{tab:channels}, and largely follow the form described in the contracted quantum learning procedure. To implement these random unitary channels efficiently, we follow the method of Monte Carlo sampling described in Ref. \cite{scottpeetz}.

\section{Results}\label{sec:results}

The results above indicate that the individual optimization procedures quickly converge and possess certain scalable features. There also is substantial literature related to these methods, as they are important routines for several algorithms and protocols. However, in the multi-objective procedure, we do not find consistent behaviors, with convergence and accuracy heavily dependent on learning rates and how we treat the objectives. There also is substantial freedom concerning the choice of unitary and the Pauli vector in the system. For channels that are close to the identity, we often find the potentially expected solution, but further from the identity, multiple solutions can usually be found, particularly due to the stochastic nature of the algorithm.

In Figure~\ref{fig:two_stage} we show varying two-stage optimization sequences where both the unitary and Pauli degrees of freedom are treated separately. In the leftmost column, the Pauli optimization is carried out first; in the middle, the unitary optimization is carried out first, and in the right, we alternate the Pauli and unitary optimizations. The target channel here has coefficients $p_x = p_z = 0.05$, and $p_y=0.3$, with $U_f = \exp \frac{i}{\sqrt{4}}\sigma_x$ and $V_f = \exp -\frac{i}{\sqrt{4}}\sigma_z$. The black points represent local unitarily equivalent solutions of the channel.

When the Pauli optimization is carried out first, the obtained red points represent the Pauli twirled channel, i.e. the diagonal of the Pauli transfer matrix when rotated by the unitary transformations. Relaxing the optimization via unitary transformations allows us to improve the optimization, but depending on the location, can sometimes yield non-optimal solutions. With no initial guess, the solution instead takes the shortest distance to the equidistant point $X=Y=Z=\frac{1-p_I}{3}$, which lies on the plane where $p_I = 0.6$. When the unitary optimization is carried out for differing initial Pauli channels, for most points this aids in biasing the solutions towards the proper direction. However, this is not always the case, and similarly, we end up on the plane of the identity instead of at properly oriented points. 

The right depicts a balanced scheme where we alternate optimizations of the Pauli and unitary channels. While we can generate varying rates of convergence and quality of solutions, in general we find that allowing for more unitary rotations compared with Pauli rotations allows for more consistent solutions (i.e. at each step we take $3$ unitary steps and 1 Pauli step). This holds even for more obscure starting configurations. We also see that multiple solutions beside $Y=0.3$ are found, i.e. where the $X$ or $Z$ channel is dominant, showing the potential freedom in the locally equivalent unitaries. This approach also yields the most accurate channels, with an order of magnitude lower trace distance than unitary or Pauli-dominated approaches. 

\begin{figure}[t]
\includegraphics[scale=1]{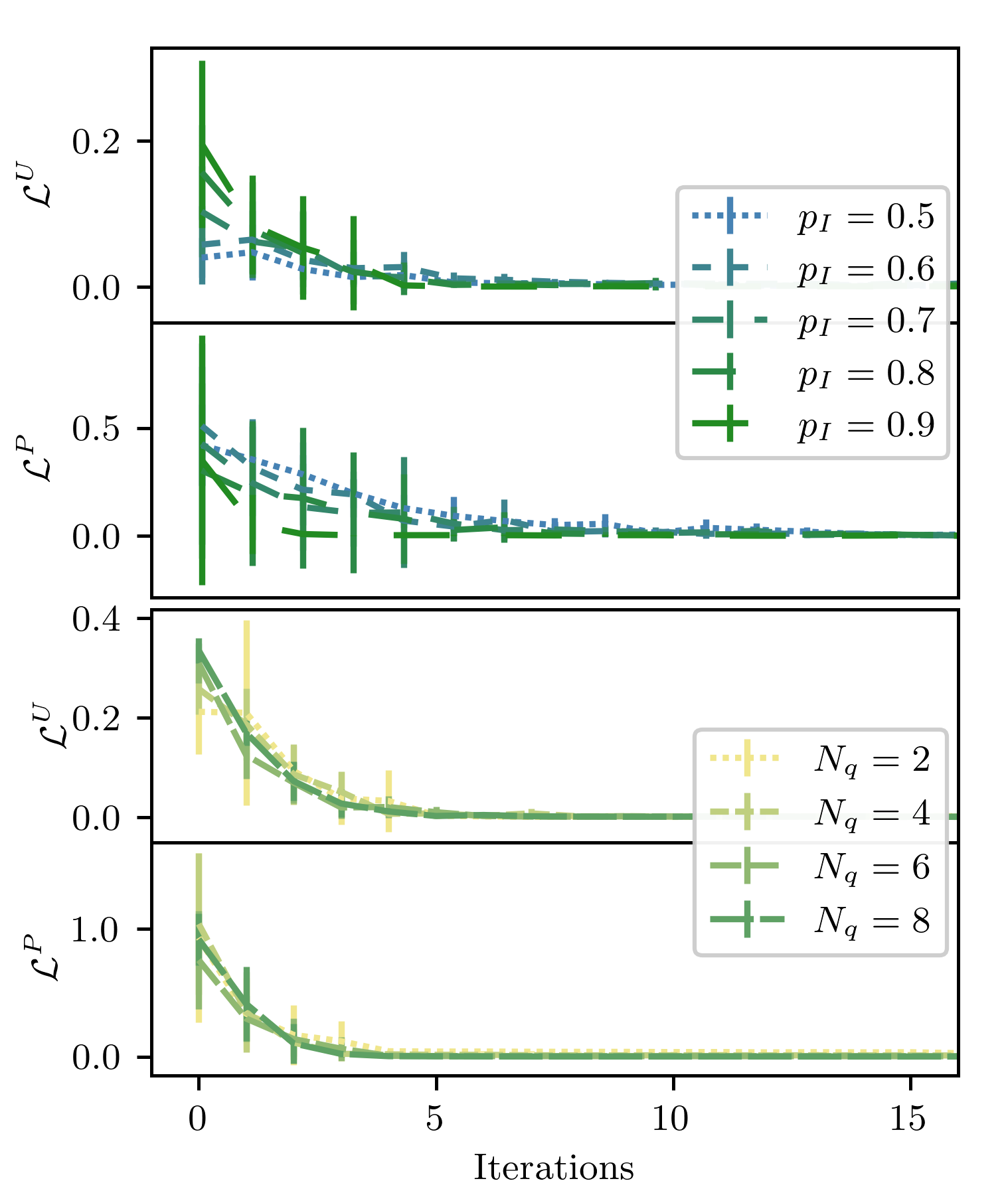}
\caption{ORUC learning for single qubit channels. (Top) Varying strengths of Pauli channel with Haar random unitary transforms. (Bottom) Given Haar random unitaries and a single additive Pauli model, we see relatively quick convergence for multiple qubits with fixed $p_I = 0.7$. Results are averaged over ten runs with varying initializations for non-identity Pauli coefficients.}
\label{fig:learn_oruc}
\end{figure}

We also show that with alternating unitary and Pauli optimizations we can perform these calculations for varying strengths in the Pauli channel, as well as scale our optimization using the additive sparse Pauli model. Figure~\ref{fig:learn_oruc} depicts these results averaged over numerous randomly initialized Pauli and unitary channels. For the varying qubit case, we use an additive model with $p_I = 0.7$ that applies the learning procedure over multiple qubits. If we were to instead use a multiplicative model for this channel, we would need higher order terms, which can be incorporated relatively easily into the model.

\begin{figure}
\includegraphics[scale=1]{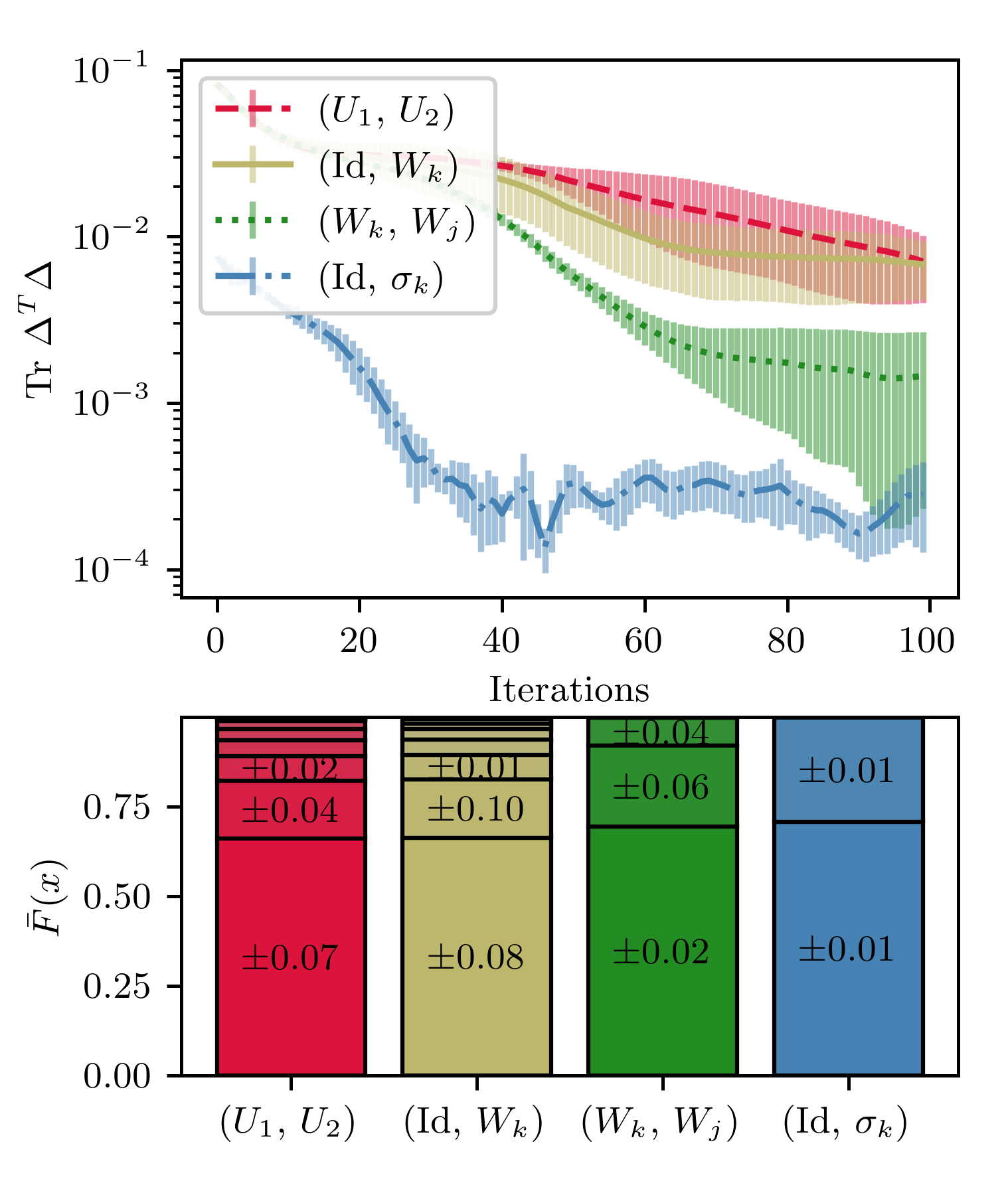}
\caption{Average channel performances for varying channels with probability weights $0.7$ and $0.3$, and pairs of unitaries. $W_k$ refers to a randomly selected non-identity Weyl channel, and $U_i$ to a Haar-random unitary channel. (Top) trace channel distance versus the number of iterations averaged over 5 runs. (Bottom) Ordered cumulative probability distributions of final channel with variance in the coefficients.}
\label{fig:weyl}
\end{figure}

Finally, we show how this procedure performs for classes of channels outside of the equivalence class. We use the set of $d=4$ Weyl matrices as a demonstration, which are a form of shift-and-multiply bases \cite{klappeneckerMonomialityNiceError2005}. These matrices are defined for arbitrary $d$ as:
\begin{equation}
W_{kj} = \sum_{m=0}^{d-1} e^{\frac{2 \pi i}{d}k m} | m+j \rangle \langle m |.
\end{equation} 
Different shift-and-multiply groups form separate equivalence classes under the unitary relation, and the Weyl channels are not locally equivalent to the Pauli channels.Figure~\ref{fig:weyl} shows the ORUC learning procedure for pairs of channels with coefficients 0.7 and 0.3, namely: random unitary channels; a Weyl and identity channel; two Weyl channels, and; an identity and Pauli channel. 

The unitary channels are Haar random, and are not necessarily orthogonal. Thus we see that the probability distribution does not exactly match the chosen unitaries. Regardless, we are able to reduce the distance by about an order of magnitude. The Weyl cases yield some interesting insights. Whereas a single unitary can be transformed into a Weyl element, two elements cannot be simultaneously transformed from the Pauli basis. Furthermore, the transformation $\sigma_k = V W_k$, does not preserve the identity, and so the identity-dominant ends up worse. While $V W_k$ is not a Weyl channel, apparently a close solution to $W_j$ is able to be found. Finally, the Pauli matrices perform the best, and reach the sampling limit for the given number of shots relatively quickly, matching the target channel almost identically.

\section{Discussion and Conclusion}
The orthogonal random unitary channel presents a class of channels that are a substantial simplification over unital and random unitary channels, but which generalize the notion of Pauli channels. By incorporating coherent effects into our channel, we allow for numerous channels to be constructed. We expect relevant applications in quantum information and in characterizing coherent noise. The trace distance here is not the common channel distance metric, which is instead the diamond distance \cite{benentiComputingDistanceQuantum2010}, though is challenging to calculate numerically. The accuracy will still be sufficient for many applications, and the scheme is scalable to larger systems. 

Despite being derived for the current iterative framework and minimizing execution time, the simplex and contracted quantum learning procedures represent tools which can be used in other contexts as well. These provide invaluable tools beyond current Pauli and unitary learning techniques, though we believe they are well suited for these approaches. There exists a slight bias in our resolution of the identity $-$ while a randomly selected Pauli does exactly resolve to the identity, we take advantage of the set of all locally commuting Pauli observables up to a given rank. For instance, in measuring $XZ$, we also can obtain information on $XI$ and $IZ$. The result is that we bias our sampled Paulis towards $k-$local portions of the identity, and improved convergence might be obtained using truly random input and output states, i.e. drawing from the Clifford group (see Appendix \ref{sec:computational}). 

In this work, we generated a learning procedure for learning random unitary channels whose elements constitute a unitary basis. From their mathematical structure, our model can only be applied to elements within a particular equivalence class, and so we must fix an error basis, which here we take to be the Pauli basis. The approaches can quickly reduce the loss function, which represents a sampled slice of the Pauli transfer matrix distance, i.e. the trace fidelity, and we demonstrate are theoretically scalable to larger systems with similar sparse structures. While we do see promising performance for more complicated channels, these will scale exponentially and we are limited by classical resource costs as well.
\vspace{0.5cm}
\section{Acknowledgments}

This work is supported by an NSF CAREER Award under Grant No. NSF-ECCS-1944085 and the NSF CNS program under Grant No. 2247007.

%%\bibliography{biblio_v2}

\appendix

\section{Computational Details}\label{sec:computational}
Here, we provide further details on the computational elements of the paper. This includes software and algorithm choices, the selection of Pauli bases and measurement of input states, and a worked example of the selection procedure. 

\textbf{Software and Optimization:} All calculations were performed on code utilizing $\textsc{qiskit}$(v0.45.x). Learning rates for stochastic gradient descent in the Pauli simplex learning were taken between 0.5 and 0.1 and for the unitary procedure at 0.1-0.05. For Figure~\ref{fig:weyl}, we used ADAM procedure for the gradient update, with standard hyperparameters \cite{kingmaAdamMethodStochastic2017}. 

\textbf{Choosing Pauli Bases Efficiently:}
While in the text we focus on the case where we sample a single Pauli input and output, this is extremely inefficient, as a standard POVM for measuring Paulis (a stabilizer state) contains information on $2^n-1$ non-identity Pauli terms. 

Assume that we have a $n-$qubit system with local qubit measurements and that we have information about our approximate channel, i.e. it is 2-local, or spatially local. We first select a single length $n$-Pauli string with no identity elements, which implies a unique measurement and input unitary from the single qubit Clifford group. One could also choose an informationally complete set of unitaries (i.e. the full Clifford group), and then randomly sample a unitary to generate a set of Pauli elements. 

Importantly, we know that the trace of any non-identity Pauli matrix is zero, so we only need to consider nonzero terms in evaluating the loss functions and gradients. Generally, the single-qubit channel case is the most straightforward, as the problem can be fully split into a tensor product space, and run completely in parallel. 

Finally, in the Pauli learning case, we choose the same configuration for both input and output and for the unitary learning, we locally choose each element to be different. This is because the Pauli channel assumes a diagonal channel in the Pauli basis. On the other hand, in the unitary gradient expressions, ${\rm Tr}~\sigma [M, \rho]$ is non-zero only if $\sigma$ and $\rho$ anticommute. 

\textbf{Sampling Paulis as Input States:}
In our scheme, a key part is the use of Paulis as input states. As these are not hermitian, we can use linearity to express the unit-norm matrix as an average over many input states. For a $n-$qubit Pauli string $\alpha$ composed of only $I$ or $Z$, we have:
\begin{align}
\sigma_\alpha &=\bigotimes_{j=1}^n \frac{1}{2}(|0\rangle \langle 0| + (-1)^{\delta^{\alpha}_Z}|1\rangle \langle 1|)  \\
&= \frac{1}{2^n} \sum_{\vec{k}} s_{\alpha}(\vec{k}) \rho(\vec k)  
\end{align}
where $\vec k$ is s bit string and $s_{\alpha}(\vec{k})$ returns the trace of the bit string against the appropriate $Z$ observable (i.e. $s_{\sigma}(\vec{k}) = (-1)^{\vec{k} \cdot \vec{z}(\sigma)}$ with $\vec{z}(\sigma)_i = \delta(\alpha_i,Z)$. Here, $\rho(\vec k)$ maps a bit to a density matrix.

Because this expression has normalization $1$, we can evaluate it at unit cost while sampling from random Pauli eigenstates. That is: 
\begin{equation}
f(\sigma_\alpha) = \frac{1}{2^n}\sum_{\vec{k}} s_\alpha(\vec k)   f( \rho(\vec{k})).
\end{equation}

In the case that our function (i.e. the channel $\E$) is separable (or at least, assumed to be separable), we can potentially simplify $\alpha$ further. That is, $ZII$, $IZI$, and $IIZ$ can be measured with only 2-state configurations, $|000\rangle \langle 000|$ and $|111\rangle \langle 111| $, due to their compatability under the trace operator.

\textbf{Example Selection Procedure:}
Here we provide an example of the selection procedure. Letting $n=3$ and assuming the channel has local nearest neighbor interactions acting on a linear graph, we first choose two random Pauli strings, $\alpha = XZY$, and $\beta = YXZ$, where we use the string notation. For $2-$local channels, we are interested in 1- and 2-local Pauli terms that form a mutually commuting set and thus can be measured or prepared simultaneously. 

In the local case, the change of basis input unitary corresponds to $U_I = S_1 H_1 H_2$, and the output unitary is $U_O = H_1 H_3 S_3^\dagger$. Thus, acting on the initial $Z$ input and output bases subject to a nearest neighbor condition yields the following possible Pauli expectation pairs:
\begin{align}
\begin{split}
    (\alpha_i,\beta_i) \in &\{XII,IZI,IIY,XZI,IZY \} \\ &\times \{ {YII},IXI,IIZ,YXI,IXZ\}.
\end{split}
\end{align}

The input states can be prepared from linear combinations of the following input states:
\begin{align}
    R = \{|0\rangle \langle 0|,|1\rangle \langle 1| \} ^{\otimes 3}
\end{align}
which when conjugated with $U_I$ provided a complete set for resolving each Pauli, and can be prepared using an $X$ gate acting on the initial $|0\rangle\langle 0|^{\otimes 3}$ state. Here, we also are abusing notation slightly to indicate tensor products of sets yielding sets of tensor product elements. $R$ is also exactly the set of POVMs, so we can quite easily prepare all possible measurement states. In this case, we would have 25 observables of interest to be reconstructed for a single set of measurement outcomes, although some of these will likely be zero based on assumptions about the channel. Specifically, the connectivity matrix (denoting potential interactions through our channel) has the form:
\begin{equation}
    \begin{matrix}
        & YII & IXI &IIZ &YXI &IXZ \\
    XII & 1   & 1   & 0  & 1  &  0 \\ 
    IZI & 1   & 1   & 1  & 1  &  1 \\ 
    IIY & 0   & 1   & 1  & 0  &  1 \\ 
    XZI & 1   & 1   & 0  & 1  &  0 \\ 
    IZY & 0   & 1   & 1  & 0  &  1 \\ 
    \end{matrix}
\end{equation}
and so we have only 17 observables which we are assuming can be coupled via nearest-neighbor interactions. This is substantially less than the full 49 non-identity terms and 63 fully connected case.

Taking the loss of a particular term $\sigma_{XZI}$ and $\rho_{IZI}$, we can express this as a linear combination of terms that can be measured:
\begin{align}
\begin{split}
        \llangle \sigma_{XII}| \E | \rho_{IXI} \rrangle = 
        \sum_{\vec k \vec j} \frac{1}{8} p({\vec j | \vec k}) s_{ZII}(\vec{k})s_{IZI}(\vec{j}) %&\llangle \rho(\vec{j})|\E_{U_O} \E \E_{U_I}| \rho(\vec{k}) \rrangle
\end{split}
\end{align}
where $ZII = U_O (XII) U_O^\dagger$ and $IZI = U_I (IZI) U_I^\dagger$, $\vec k$ and $\vec j$ are computational $Z-$basis POVMs, and $p({\vec j|\vec{k}})$ is the measured probability of the outcome $\vec j$ given input state $\rho(\vec k)$ when the channel $\E$ is applied. For instance, given 1000 shots, we can allocate 125 shots to each input state, and then reconstruct the PTM element according to how the measurement shots are distributed.

\section{Channel Distances and Loss Functions}\label{sec:app_loss}
Here we provide more details on the different loss functions, and specifically the RI CQL method. Representing the loss function (without a factor of $\frac{1}{2}$) in vectorized notation, we have:
\begin{align}
\begin{split}
   \mathcal{L}(\sigma,\rho) &= {\rm Tr}\sigma (\mathcal{F}[\rho] - \mathcal{E}[\rho])^2 \\ 
&= \llangle \sigma| \mathcal{F} - \mathcal{E} | \rho \rrangle \llangle \rho | \mathcal{F}^\dagger - \mathcal{E}^\dagger | \sigma  \rrangle \end{split}
\end{align}
By replacing $\sigma$ and $\rho$ with their averaged channels we obtain:
\begin{align}
{\mathbb{E}_{\rho,\sigma}} [L(\sigma,\rho)] &\propto {\rm Tr} ( \F - \E)(\F - \E)^\dagger. 
\end{align}

Based on the trace cyclic condition, we have:
\begin{align}
\begin{split}
    {\rm Tr} \mathcal{F} \mathcal{E}^\dagger &= {\rm Tr} \mathcal{F} \mathcal{E}_V^\dagger \mathcal{E}_P \mathcal{E}_U^\dagger \\
&= \mathbb{E}_{\sigma, \rho} \llangle \sigma |\mathcal{E}_U^\dagger \mathcal{F} \mathcal{E}_V^\dagger | \rho \rrangle \llangle \rho | \mathcal{E}_P | \sigma \rrangle  \end{split}
\end{align}
which matches the Pauli learning procedure. Resolving the identity with respect to the unitary channels yields:
\begin{align}
\mathcal{L}_A'(\sigma, \rho) &= \frac{1}{2} |\llangle \sigma | \E_V^\dagger \E_P  \E_U^\dagger \F - \E_A| \rho \rrangle|^2, \\
\mathcal{L}_B'(\sigma, \rho) &= \frac{1}{2} |\llangle \sigma | \F \E_V^\dagger \E_P \E_U^\dagger - \E_B| \rho \rrangle|^2 .
\end{align}

These terms differ from the given loss only by the square of the channel evaluations. It is straightforward to show that these square terms do not contribute to the gradients (i.e. $\E_{A/B}$ are unitary), and so the gradient terms are the same. 

These loss functions however can have non-zero minimum due to the differences in the square of the channel evaluations. Specifically, we have:
\begin{equation}
\mathbb{E}[\mathcal{L}_A' - \mathcal{L}] = {\rm Tr}\big( \E^\dagger \F \F^\dagger \E + \mathbb{I} - \F \F^\dagger - \E \E^\dagger \big). 
\end{equation}
As these are potentially non-zero, using the gradient norm in place of the loss is particularly useful. Generally, for the ORUC channels we are looking at here, the difference in these functions comes mostly from the same Pauli elements, i.e. when for $\sigma \equiv \rho$. Inputting $\sigma ~\equiv \rho$ will generally lead to a $0$ in the loss function. At the same time, regardless of the loss function, elements where $\sigma ~\equiv \rho$ do not contribute directly to the gradient of $A$ or $B$, due to the commutator expression only being satisfied for identity terms, indicating these contribute to a global phase. Thus, these can be largely omitted and the loss function maintains a zero minima. The explicit forms of the channels we use here are given in Table \ref{tab:channels}, and we give the corresponding algorithm for this approach in Alg. \ref{alg:cqeroi}.

\begin{table}[h]
\caption{Different channel forms for loss functions related to orthogonal random unitary channels.}

\begin{tabular}{c|cc}
Loss Functions & Trial Channel & Target Channel \\
\hline 
Generic & $\E_U \E_P \E_V$ & $\F$ \\
Pauli & $\E_P$ & $\E_U^\dagger \F \E_V^\dagger$ \\ 
\hline
Input Unitary & $\E_A $  &  $ \E_V^\dagger \E_P  \E_U^\dagger \F$  \\
Output Unitary & $\E_B$ & $\F \E_V^\dagger \E_P \E_U^\dagger $ \\
\end{tabular}
\label{tab:channels}
\end{table}

\begin{algorithm}[H]
\caption{Resolution-of-the-identity contracted quantum learning algorithm for learning the outer and inner unitary channels $U,V$. Given: A target channel $\F$, learning rate $\eta$, number of iterations $N^U$, current guesses $U,V$, indices $\mathcal{I}_\mathcal{A}$ of a set of generators $\mathcal{A}$, and Pauli channel $\E_P$}
\label{alg:cqeroi}
\begin{algorithmic}[1]
\State \textbf{inputs}: $\F,\mu,N^U,\E_U,\E_V,\E_P$
\State \textbf{sample}: $[\sigma_1,...,\sigma_{N^U}], [\rho_1,...,\rho_{N^U}] \subset\mathbb P^n, \mathbb P^n$
%\State \textbf{sample}: $[\rho_1,...,\rho_{N^U}]$
\For{{$\alpha=1,...,N^U$}}
    \State \textbf{measure} $y_\alpha^A=\text{Tr}\big[\sigma_\alpha \E^\dagger_V \E_P \E_U^\dagger \F[ \rho_\alpha ]\big]$
    \State \textbf{measure} $y_\alpha^B=\text{Tr}\big[\sigma_\alpha \F \E_V^\dagger \E_P \E_U^\dagger [ \rho_\alpha ]\big]$
    \State \textbf{calculate} $x_\alpha= \delta(\sigma_\alpha, \rho_\alpha)$
    \For{{$k \in \mathcal{I}_\mathcal{A}$}}
        \State \textbf{calculate} $\frac{\partial }{\partial a_k / b_k} x_\alpha(t) |_{t=0} = i \delta(\sigma_\alpha, [\sigma_k,\rho_\alpha])$
    \EndFor
    \State \textbf{initialize} $A=-\sum_k i[y_\alpha^A - x_\alpha]\frac{\partial}{\partial a_k}x(t)|_{t=0}~\sigma_k$
    \State \textbf{initialize} $B=-\sum_k i[y_\alpha^B - x_\alpha]\frac{\partial}{\partial b_k}x(t)|_{t=0}~\sigma_k$
    \State $\mathcal{E}_V \gets \mathcal{E}_V \circ \mathcal{E}_{\exp \eta A'}$
    \State $\mathcal{E}_U \gets \mathcal{E}_{\exp \eta B'} \circ \mathcal{E}_U$
\EndFor
\State \textbf{return} $\mathcal{E}_U,\mathcal{E}_V$
\end{algorithmic}
\end{algorithm}

\section{Contractions of the Least-Squares Problem}\label{sec:con_lstsq}
Consider the least square problem of learning a state $|\psi\rangle$ from an initial state $|\phi\rangle $:
\begin{equation}
   \mathcal{L}[|\phi\rangle] = \frac{1}{2}||~|\phi\rangle - |\psi \rangle~ ||^2
\end{equation}
If we consider the variation of the loss with respect to $|\phi\rangle$, we have:
\begin{equation}
    \frac{d \mathcal{L}}{d\phi}  = \frac{1}{2}\big( \langle \delta \phi |\psi \rangle + \langle \psi |\delta \phi \rangle - \langle \delta\phi| \phi \rangle - \langle \phi | \delta\phi\rangle). 
\end{equation}
When this residual expression is 0, we have reached a local minimum of the problem with respect to our variation. In general, we can take sets of variations, and then look for a solution which minimizes all variations. 

For the least-squares problem, if the set of variations is informationally complete, our states are equivalent. By contradiction, for two non-equal states, some residual from an informationally complete set will be non-zero. Regardless, here our variations are akin to operators acting locally on a bipartite state (through vectorization of $\rho$):
\begin{align}
|\delta \phi^A_j \rangle = \sum_k \sigma_j \otimes I | \alpha_k\rangle \otimes | \beta_k \rangle   \\
|\delta \phi^B_j \rangle = \sum_k I \otimes \sigma_j  | \alpha_k\rangle \otimes | \beta_k \rangle   
\end{align}
This is not informationally complete with respect to the entire Hilbert space, but is over the space of local unitaries. By comparison, for a stationary point of a variational ansatz, we have
\begin{equation}
    \frac{dL}{d\theta} = \frac{dL}{d\phi}\frac{d\phi}{d \theta} = 0
\end{equation}
which also considers the dependence of $\psi$ on $\theta$. The loss is stationary with respect to variations in $\theta$ but not all possible variations of the wavefunction $\phi$. 

\section{On Sparse Additive Channels}\label{sec:add}
Here, we show that for certain classes of noise models and target observables, the effects of a multiplicative Pauli channel, which produces higher order Pauli terms, can be described by a low-order additive channel. 

Consider a set of $k-$local Pauli operators $A = \{\sigma_i\}$, as well as a $k-$local observable $\hat{O}$, composed of elements of $A$. Then, we consider a channel generated from iterative $k-$local operators, each with a probability $1 \geq p_i \geq 0$, $\{\mathcal{E}_i\}$ defined as:
\begin{equation}
\mathcal{E}_i[\rho] = (1-p_i)\rho + p_i  \sigma_i \rho \sigma_i 
\end{equation}
The global channel then has the form:
\begin{align}
\mathcal{F}[\rho]  = \mathcal{E}_0 \circ \mathcal{E}_1 \circ \cdot \cdot \cdot \mathcal{E}_n [\rho] . 
\end{align}

Consider the elements of the Pauli transfer matrix for the product and a single additive  channel $\mathcal{G}$, we have:
\begin{align}
f_\alpha &= \frac{1}{d}{\rm Tr}~ \sigma_\alpha \mathcal{F}[\sigma_\alpha] \\ &= \prod_{k}(1-q_k +  S_{k\alpha} q_k)	\\  
g_\alpha &= \frac{1}{d}{\rm Tr}~ \sigma_\alpha \mathcal{G}[\sigma_\alpha] \\
&= p_I + \sum_{k} S_{k\alpha} p_k 
\end{align}
If we consider the quantities on average, we can simplify some of the expressions. Let $\Delta$ be the average difference in the number of commuting ($N_c$) and anti-commuting ($N_a$) terms (where for sparse channels on multiple qubits, generally $\Delta >0$ (see Table \ref{tab:Delta}). Then we have that the dimension of the Pauli channel $d = N_a + N_c$, $\Delta = N_c - N_a$, and $N_a = \frac{d - \Delta}{2}$. 

\begin{figure*}
\includegraphics[scale=0.65]{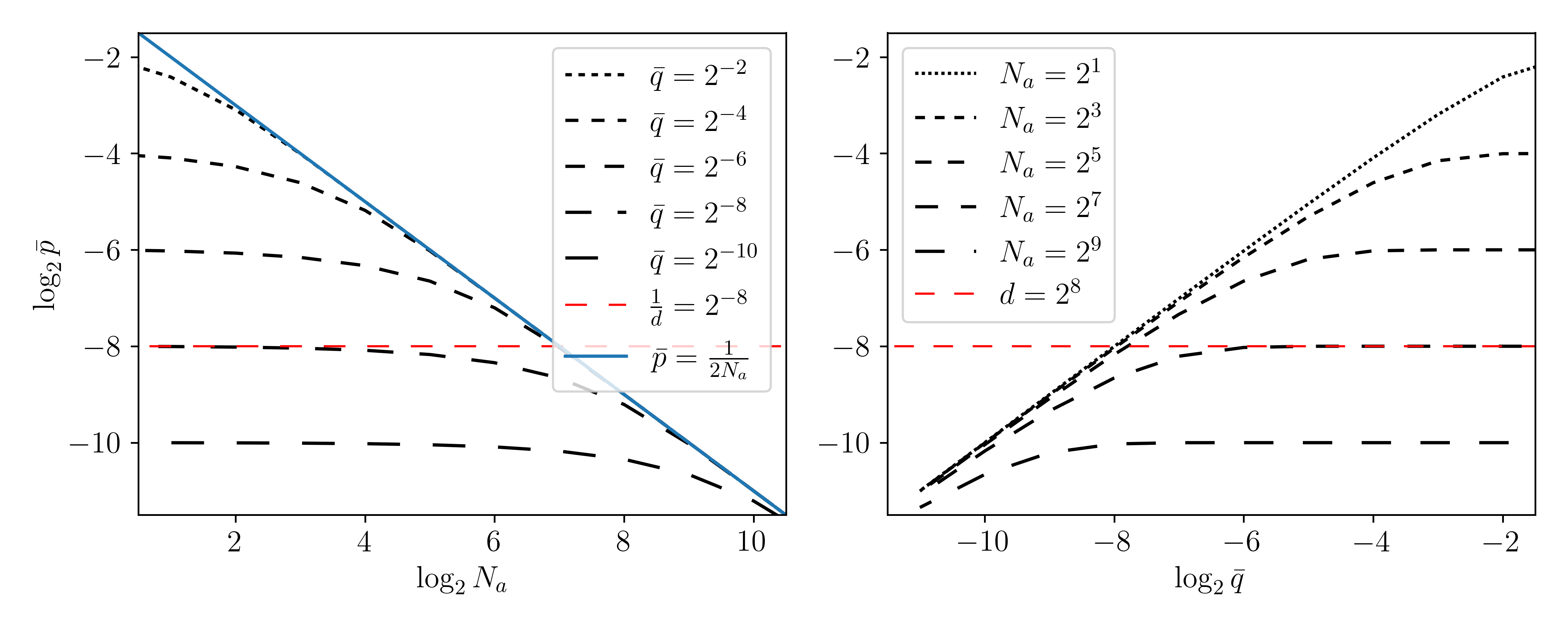}
\label{fig:a1}
\caption{Given an average probability under a multiplicative model $\bar{q}$, with varying number of anticommuting elements $N_a$, we show the resulting average probability required for the additive model to model the multiplicative one, as a function of the number of anticommuting terms and the average probability.}
\end{figure*}

\begin{table}
\caption{Difference in the number of commuting and anti-commuting terms generated by sparse sets of indexed sets. Also given are the numerical values of $\Delta$ for the first few numbers of qubits. Generally,for $k-$local sets we observe a positive (commuting-favored) polynomial $\Delta$.}
\begin{tabular}{c|c|c}
Layout & $\langle\Delta \rangle $ & $\Delta_4, \Delta_6, \Delta_8, \Delta_{10}, \Delta_{12}$ \\ \hline
$\{(i)\}$ & $3N-4$ &  $8, 14, 20, 26, 32 $  \\
$\{(i,i+1)\}$ & $11.6 N - 38.6$ & $10.0, 31.0, 53.6, 76.8, 100.33$ \\
$\{(i,j)\}$ & $O(4.5 N^2)$ & $6.4, 122, 233, 380, 564$ \\
$\{(i,j,k)\}$ & $O(4.4 N^3 - 59.6 N^2)$  & $-0.07, 2.5, 158, 674 , -$ \\ 
\end{tabular}
\label{tab:Delta}
\end{table}

For the multiplicative channel, we can replace $q_k$ with the average $\bar{q}$ and consider the average over $\alpha$:
\begin{align}
\langle f_\alpha \rangle &= \prod_k (1- \bar{q} + S_{k\alpha} \bar{q}) \\
&= ( 1- 2\bar{q})^{N_a}
\end{align}
Generally, $\bar{q} \ll \frac{1}{2}$, implying that $f_\alpha > 0$. For the additive channel we have:
\begin{align}
\langle g_\alpha \rangle &= 1- d \bar{p} + \sum_{k} S_{k\alpha} \bar{p} \\
&= 1 - d \bar{p} + \Delta \bar{p} \\
&= 1 + (\Delta - d) \bar{p} \\
&= 1 - 2 N_a \bar{p}
\end{align}

Additionally, this implies an equivalence condition between the two average fidelities when:
\begin{equation}
\bar{p} = \frac{1 - e ^{\bar{\lambda} N_a} }{2 N_a}
\end{equation}
where $\bar{\lambda} = \ln (1-2 \bar{q})$ 

We can derive some constraints based on the positivity of $g_\alpha$ as well as the normalization of the average probability, we have:
\begin{align}
\bar{p} &\leq \frac{1}{2 N_a}, \\
\bar{p} &\leq \frac{1}{d}.
\end{align}

For many of our sparse models, $2 N_a \ll d$, implying that the second constraint on $\bar p$ is stronger. This results in the following constraint on $\bar p$:
\begin{align}
\frac{1 - e ^{\bar{\lambda} N_a} }{2 N_a} &\leq \frac{1}{d} \\
\rightarrow 1 - e^{\bar{\lambda} N_a} &\leq \frac{2N_a}{d}  \\
\rightarrow e^{\bar{\lambda} N_a}  &\geq \frac{d  - 2 N_a}{d} \\
 &= \frac{\Delta}{d}
\end{align}
In general, this sets a condition on the value of $\bar{q}$ which can be represented in this way. We note that to first order this gives $\bar{q} \leq \frac{1}{d}$, which holds for a variety of modern quantum systems. However, the actual limit depends more on the right-hand term, i.e. the ratio of the difference in commuting elements to the dimension of the system. 

We show the possible $\bar{p}$ given values of $\bar{\lambda}$ and varying number of anticommuting elements in Figure ~\ref{fig:a1}. Here, the red line suggests a potential boundary of failure in the additive model. Given a very large number of terms, i.e. $d=2^8$, we can vary $N_a$ up to $2^7$. On the left, we see that small $\bar{p}$ requires that $\bar{q}$ be sufficiently small as well regardless of the $N_a$. On the right figure, we also see that for $\bar{q} \geq 2^{-8}$ with $d=2^8$, acceptable values of $\bar{p}$ do exist, though $N_a$ is required to be much smaller.

\end{document}